\documentclass[a4paper,11pt]{amsart}
\usepackage{amssymb}
\usepackage{enumerate}
\def \Z {\mathbb Z}
\def \R {\mathbb R}
\def \C {\mathbb C}

\def \N {\mathbb N}

\def\dd{\mathrm d}
\def\ee{\mathrm e}
\def\ii{\mathrm i}
\DeclareMathOperator{\AB}{AB}
\DeclareMathOperator{\Arcsin}{Arcsin}
\DeclareMathOperator{\Arg}{Arg}
\DeclareMathOperator{\Ln}{Ln}

\DeclareMathOperator{\supp}{supp}
\theoremstyle{plain}
\newtheorem{thm}{Theorem}
\newtheorem{prop}{Proposition}
\newtheorem{cor}{Corollary}
\theoremstyle{definition}
\newtheorem{Def}{Definition}
\theoremstyle{remark}
\newtheorem{rem}{Remark}
\begin{document}
\title[Continuous matrix-valued Anderson model]
       {Positivity of Lyapunov exponents for a 
          continuous matrix-valued Anderson model}
\author[H.~Boumaza]{Hakim Boumaza}
\address{Institut de Math\'ematiques de Jussieu\\ 
Universit\'e Paris 7 Denis Diderot \\ 
2 place Jussieu,
75251 Paris, France}
\email{boumaza@math.jussieu.fr}
\date{March 20, 2007}
\begin{abstract}
We study a continuous matrix-valued Anderson-type model. Both leading Lyapunov exponents of this model are proved to be positive and distinct for all energies in $(2,+\infty)$ except those in a discrete set, which leads to absence of absolutely continuous spectrum in $(2,+\infty)$. This result is an improvement 
of a previous result with Stolz. The methods, based upon a result by Breuillard and Gelander on dense
subgroups in semisimple Lie groups, and a criterion by Goldsheid and Margulis, allow for singular Bernoulli distributions.
\end{abstract}
\maketitle
\section{Introduction}

We will study the question of separability of Lyapunov exponents for a continuous matrix-valued Anderson-Bernoulli model of the form: 
\begin{align}\label{modelAB2}
H_{\AB}(\omega)&=-\frac{\dd^{2}}{\dd x^{2}}+ \begin{pmatrix}
0 & 1 \\
1 & 0
\end{pmatrix}\nonumber\\
&\quad
+ \sum_{n\in \Z}
\begin{pmatrix}
\omega_{1}^{(n)} \chi_{[0,1]}(x-n)& 0 \\
0 & \omega_2^{(n)} \chi_{[0,1]}(x-n)
\end{pmatrix}  
\end{align}
acting on $L^{2}(\R)\otimes\C^{2}$. This question is coming from a more general problem on Anderson-Bernoulli models. Indeed, localization for Anderson models in dimension $d\geq 2$ is still an open problem if one look for arbitrary disorder, especially for Bernoulli randomness. A possible approach to try to understand localization for $d=2$ is to discretize one direction. It leads to consider one-dimensional continuous Schr\"odinger operators, no longer scalar-valued, but now $N\times N$ matrix-valued. Before trying to understand how to handle with $N\times N$ matrix-valued  continuous Schr\"odinger operators, we start with the model (\ref{modelAB2}) corresponding to $N=2$.

What is already well understood is the case of dimension one scalar-valued continuous Schr\"odinger operators with arbitrary randomness including Bernoulli distributions (see \cite{stolz2}) and discrete matrix-valued Schr\"odinger operators also including the Bernoulli case (see \cite{goldsheid} and \cite{klein}). We aim at combining existing techniques for these cases to prove that for our model (\ref{modelAB2}), the Lyapunov exponents are all positive and distinct for all energies outside a discrete set, at least for energies in $(2,+\infty)$ (see Theorem \ref{HAB2thm}).

It is already proved in \cite{stolzboumaza} that for model (\ref{modelAB2}), the Lyapunov exponents are separable for all energies except those in a countable set, the critical energies. But the techniques used in \cite{stolzboumaza} didn't allow us to avoid the case of an everywhere dense countable set of critical energies. Due to Kotani's theory (see \cite{kotanis}) this result will imply absence of absolutely continuous spectrum in the interval $(2,+\infty)$. But we keep in mind that we want to be able to use our result to prove Anderson localization and not only the absence of absolutely continuous spectrum. The separability of Lyapunov exponents can be view as a first step  in order to follow a multiscale analysis scheme. The next step would be to prove some regularity on the integrated density of states, like local H\"older-continuity and then to prove a Wegner estimate and an Initial Length Scale estimate to start the multiscale analysis (see \cite{stollmann}). To prove the local H\"older-continuity of the integrated density of states, we need to have the separability of the Lyapunov exponents on intervals (see \cite{carmona} or \cite{stolz2}). But, if like in \cite{stolzboumaza} we can get an everywhere dense countable set of critical energies, we will not be able to prove local H\"older-continuity of the integrated density of states. That is why we need to improve the result of \cite{stolzboumaza}.

Our approach of the separability of Lyapunov exponents is based upon an abstract criterion in terms of the group generated by the random transfer matrices. This criterion has been provided by
Gol'dsheid and Margulis in \cite{goldsheid}. It is exactly this criterion which allowed to prove Anderson localization for discrete strips (see \cite{klein}). This criterion is also interesting because it allows for singularly distributed random parameters, including Bernoulli distributions. 

We had the same approach in \cite{stolzboumaza}; what changes here is the way to apply the criterion of Gol'dsheid and Margulis. To apply this criterion we have to prove that a certain group is Zariski-dense in the symplectic group 
$\mathrm{Sp}_2(\R)$. In \cite{stolzboumaza} we were constructing explicitly a family of ten matrices linearly independent in the Lie algebra $\mathfrak{sp}_2(\R)$ of $\mathrm{Sp}_2(\R)$. This construction was only possible by considering an everywhere dense countable set of critical energies. By using a result of group theory by Breuillard and Gelander (see \cite{breuillarda}), we are here able to prove that the group involved in Gol'dsheid-Margulis' criterion is dense in $\mathrm{Sp}_2(\R)$ for all energies in $(2,+\infty)$, except those in a discrete subset.

We start at Section \ref{sec:separability} with a presentation of the necessary background on products of i.i.d.\ symplectic matrices and with a statement of the criterion of Gol'dsheid and Margulis. We also present the result of Breuillard and Gelander in this section. Then, in Section \ref{secmodel} we make precise the assumptions made on the model (\ref{modelAB2}) and we explicit the transfer matrices associated to this model. In Section \ref{secproof} we give the proof of our main result, Theorem \ref{HAB2thm} by following the steps given by the assumptions of Theorem \ref{breuillard} by Breuillard and Gelander.

We finish by mentioning that different methods have been used to prove localization properties for random operators on strips in \cite{KMPV}. They are based upon the use of spectral averaging techniques which did not allow to handle with singular distributions of the random parameters. So even if the methods used in \cite{KMPV} (which only considers discrete strips) have potential to be applicable to continuous models, one difference between these methods and the ones used here is that, like in \cite{stolzboumaza}, we handle singular distributions, in particular Bernoulli distributions.

\section{Criterion of separability of Lyapunov exponents}
\label{sec:separability}

We will first review some results about Lyapunov exponents and how to prove their separability. These results hold for general sequences of i.i.d.\ random symplectic matrices. Even if we will only use them for symplectic matrices in $\mathcal{M}_{4}(\R)$, we will write these results for symplectic matrices in $\mathcal{M}_{2N}(\R)$ for arbitrary $N$.

Let $N$ be a positive integer. Let $\mathrm{Sp}_{N}(\R)$ denote the
group of $2N\times 2N$ real symplectic matrices, i.e.:
\[
\mathrm{Sp}_{N}(\R)=\{M\in\mathrm{GL}_{2N}(\R)\mid {}^tMJM=J\}
\]
where 
\[
J=
\begin{pmatrix}
0 & -I \\
I & 0
\end{pmatrix}.
\] 
Here, $I=I_N$ is the $N\times N$ identity matrix.

\begin{Def}[Lyapunov exponents]\rm
Let $(A_{n}^{\omega})_{n\in \N}$ be a sequence of i.i.d.\ random
matrices in $\mathrm{Sp}_{N}(\R)$ with 
\[
\mathbb{E}(\log^{+}
||A_{1}^{\omega}||) <\infty.
\] 
The \emph{Lyapunov exponents}
$\gamma_{1},\ldots,\gamma_{2N}$ associated with
$(A_{n}^{\omega})_{n\in \N}$ are defined inductively by
\[
\sum_{i=1}^{p} \gamma_{i} = \lim_{n \to \infty} \frac{1}{n}
\mathbb{E}(\log ||\wedge^{p} (A_{n}^{\omega}\ldots A_{1}^{\omega})
||).
\]
\end{Def}

\noindent
Here $\wedge^{p} (A_{n}^{\omega}\ldots A_{1}^{\omega})$ denotes the $p$-th exterior power of the matrix 
$(A_{n}^{\omega}\ldots A_{1}^{\omega})$, acting on the $p$-th exterior power of $\R^{2N}$. 
For more details about these  $p$-th exterior powers, see \cite{bougerol}.

One has $\gamma_{1} \geq \ldots \geq \gamma_{2N}$. Moreover,
the random matrices $(A_{n})_{n\in \N}$ being symplectic, we have the
symmetry property $\gamma_{2N-i+1}=-\gamma_{i}$, for
$i=1,\ldots,N$ (see \cite{bougerol}, Proposition 3.2). 

We say that the Lyapunov exponents of a sequence $(A_{n}^{\omega})_{n\in \N}$ of i.i.d.\ random matrices are \emph{separable} when they are all distinct: 
\[
\gamma_{1} > \gamma_2 > \ldots > \gamma_{2N}.
\]
We can now give a criterion of separability of the Lyapunov exponents. For the definitions of $L_p$-strong irreducibility and $p$-contractivity we refer to \cite{bougerol}, Definitions
A.IV.3.3 and A.IV.1.1, respectively.

Let $\mu$ be a probability measure on $\mathrm{Sp}_{N}(\R)$. We denote by $G_{\mu}$ the smallest closed subgroup of $\mathrm{Sp}_{N}(\R)$ which contains the topological support of $\mu$, $\supp\mu$. 

Now we can set forth the main result on separability of Lyapunov exponents, which is a generalization of Furstenberg's theorem to the case $N>1$.

\begin{prop}\label{lyapsepmain}
Let $(A_{n}^{\omega})_{n\in \N}$ be a sequence of i.i.d.\ random
symplectic matrices of order $2N$ and $p$ be an integer, $1\leq p\leq N$. Let $\mu$ be 
the common distribution of the $A_{n}^{\omega}$. If 
\begin{enumerate}[\rm(a)]
\item
$G_{\mu}$ is $p$-contracting and $L_p$-strongly irreducible,
\item
$\mathbb{E}(\log\Vert A_{1}^{\omega}\Vert)<\infty$, 
\end{enumerate}
then the following holds:
\begin{enumerate}[\rm(i)]
\item
$\gamma_{p} > \gamma_{p+1}$
\item
For any non zero $x$ in $L_p$:
\[
\lim_{n\to \infty} \frac{1}{n}\mathbb{E}\bigl(\log\Vert\wedge^{p} A_{n}^{\omega}\ldots A_{1}^{\omega}x\Vert\bigr)=\sum_{i=1}^{p} \gamma_i\,.
\]
\end{enumerate}
\end{prop}


\begin{proof}
See \cite{bougerol}, Proposition 3.4.
\end{proof}

\begin{cor}\label{lyapsepcor}
If 
\begin{enumerate}[\rm(a)]
\item
$G_{\mu}$ is $p$-contracting and $L_p$-strongly irreducible for $p=1,\ldots ,N$,
\item
$\mathbb{E}(\log\Vert A_{1}^{\omega}\Vert)<\infty$, 
\end{enumerate}
then 
$\gamma_{1} > \gamma_2 > \ldots > \gamma_{N} > 0$.
\end{cor}

\begin{proof}
Use Proposition \ref{lyapsepmain} and the symmetry property of Lyapunov exponents. 
\end{proof}

For explicit models like (\ref{modelAB2}), it can be quite difficult to check the $p$-contractivity and the $L_{p}$-strong irreducibility for all $p$. To avoid this difficulty, we will use the Gol'dsheid-Margulis theory presented in 
\cite{goldsheid} which gives an algebraic criterion to check these assumptions. The idea is the following: if the group $G_{\mu}$ is large enough in an algebraic sense then $G_{\mu}$ is $p$-contractive and $L_{p}$-strongly irreducible for all $p$.

We first recall the definition of the Zariski topology on $\mathcal{M}_{\mathrm{2N}}(\R)$. We identify $\mathcal{M}_{\mathrm{2N}}(\R)$ to $\R^{(2N)^{2}}$ by viewing a matrix as the list of its entries. Then for $S\subset \R[X_{1},\ldots ,X_{(2N)^{2}}]$, we set: 
\[
V(S)=\{ x\in \R^{(2N)^{2}}\mid\forall P\in S,\ P(x)=0 \}
\]
So, $V(S)$ is the set of common zeros of the polynomials of $S$. These sets $V(S)$ are the closed sets of the Zariski topology on $\R^{(2N)^{2}}$. Then, on any subset of $\mathcal{M}_{\mathrm{2N}}(\R)$ we can define the Zariski topology as the topology induced by the Zariski topology on $\mathcal{M}_{\mathrm{2N}}(\R)$. In particular we define in this way the Zariski topology on $\mathrm{Sp}_{N}(\R)$. 

We can now define the Zariski closure of a subset $G$ of $\mathrm{Sp}_{N}(\R)$. It is the smallest closed subset for the Zariski topology that contains $G$. We denote it by $\mathrm{Cl_{Z}}(G)$. In other words, if $G$ is a subset of $\mathrm{Sp}_{N}(\R)$, its Zariski closure $\mathrm{Cl_{Z}}(G)$ is the set of zeros of polynomials vanishing on $G$. A subset $G'\subset G$ is said to be Zariski-dense in $G$ if $\mathrm{Cl_{Z}}(G')=\mathrm{Cl_{Z}}(G)$, i.e., each polynomial vanishing on $G'$ vanishes on $G$. 

Being Zariski-dense is the meaning of being large enough for a subgroup of $\mathrm{Sp}_{N}(\R)$ to be $p$-contractive and $L_{p}$-strongly irreducible for all $p$. More precisely, from the results of Gol'dsheid and Margulis one gets: 

\begin{thm}[Gol'dsheid-Margulis criterion, \cite{goldsheid}]\label{algthm}
If $G_{\mu}$ is Zariski den\-se in $\mathrm{Sp}_{N}(\R)$, then for all
$p$, $G_{\mu}$ is $p$-contractive and $L_{p}$-strongly irreducible.
\end{thm}

\begin{proof}
It is explained in \cite{stolzboumaza} how to get that criterion from the results of Gol'dhseid and Margulis stated in \cite{goldsheid}.
\end{proof}

As we can see in \cite{stolzboumaza}, it is not easy to check directly that the group $G_{\mu_{E}}$ introduced there is Zariski-dense. In fact, in \cite{stolzboumaza} we were reconstructing explicitly the Zariski closure of $G_{\mu_{E}}$. But this construction was possible only for energies not in a dense countable subset of $\R$. We will now give a way to prove more systematically the Zariski-density of a subgroup of $\mathrm{Sp}_{N}(\R)$. It is based on the following result of Breuillard and Gelander: 

\begin{thm}[Breuillard, Gelander \cite{breuillarda}]\label{breuillard}
Let $G$ be a real, connected, semisimple Lie group, whose Lie algebra is $\mathfrak{g}$. 

Then there is a neighborhood $\mathcal{O}$ of $1$ in $G$, on which $\log=\exp^{-1}$ is a well defined diffeomorphism, such that $g_{1},\ldots,g_{m}\in \mathcal{O}$ generate a dense subgroup whenever 
$\log g_{1},\ldots,\log g_{m}$ generate $\mathfrak{g}$.
\end{thm}

We will use this theorem in the sequel to prove that the subgroup generated by the transfer matrices associated to our operator is dense, hence Zariski-dense, in $\mathrm{Sp}_{N}(\R)$. 

In the next section we will make precise the assumptions on model (\ref{modelAB2}) and give the statement of our main result.

\section{A matrix-valued continuous Anderson model}\label{secmodel}

Let
\begin{equation}\label{model2}
H_{\AB}(\omega)=-\frac{\dd^2}{\dd x^2} + V_0 + \sum_{n\in \Z} 
\begin{pmatrix}
\omega_{1}^{(n)} \chi_{[0,1]}(x-n) & 0 \\
0 & \omega_2^{(n)} \chi_{[0,1]}(x-n)
\end{pmatrix}
\end{equation}
be a random Schr\"odinger operator acting in $L^{2}(\R)\otimes\C^{2}$. Here 
\begin{enumerate}[$\bullet$]
\item
$\chi_{[0,1]}$ denotes the characteristic function of the interval $[0,1]$, 
\item
$V_0$ is the constant-coefficient
multiplication operator by 
$\begin{pmatrix} 
0 & 1 \\ 1 & 0
\end{pmatrix}$,
\item
$(\omega_{1}^{(n)})_{n\in\Z}$, $(\omega_2^{(n)})_{n\in\Z}$ are
two independent sequences of i.i.d.\ random variables with common distribution $\nu$ such that
$\{0,1\} \subset\supp\nu$.
\end{enumerate}
This operator is a bounded perturbation of $-\frac{\dd^2}{\dd x^2}
\oplus-\frac{\dd^2}{\dd x^2}$. Thus it is self-adjoint on the Sobolev space $H^{2}(\R)\otimes\C^{2}$.

For the operator $H_{\AB}(\omega)$ defined by (\ref{model2}) we have the following result: 

\begin{thm}\label{HAB2thm}
Let $\gamma_{1}(E)$ and $\gamma_2(E)$ be the positive Lyapunov exponents associated to $H_{\AB}(\omega)$. 

There exists a discrete set $\mathcal{S}_{\mathrm{B}}\subset \R$ such that for all $E>2$, 
$E\notin\mathcal{S}_{\mathrm{B}}$, $\gamma_{1}(E)> \gamma_2(E) >0$. \end{thm}

\begin{cor}\label{HAB2cor}
$H_{\AB}(\omega)$ has no absolutely continuous spectrum in the interval $(2,+\infty)$.
\end{cor}

We will first specify some notations. We consider the differential system: 
\begin{equation}\label{system2order}
H_{\AB}u=Eu,\quad E\in \R.
\end{equation}
For a solution $u=(u_{1},u_2)$ of this system we define the transfer matrix $A_{n}^{\omega^{(n)}}(E)$, $n\in \Z$ from $n$ to $n+1$ by the relation
\[
\begin{pmatrix}
u_{1}(n+1) \\
u_2(n+1) \\
u_{1}'(n+1) \\
u_2'(n+1)
\end{pmatrix}
= A_{n}^{\omega^{(n)}}(E)
\begin{pmatrix}
u_{1}(n) \\
u_2(n) \\
u_{1}'(n) \\
u_2'(n)
\end{pmatrix}.
\]
The sequence $\{A_{n}^{\omega^{(n)}}(E)\}_{n\in \Z}$ is a sequence of i.i.d.\ random matrices in the symplectic group $\mathrm{Sp}_2(\R)$. This sequence will determine the Lyapunov exponents at energy $E$. In order to use Proposition \ref{lyapsepmain}, it is necessary to define a measure on $\mathrm{Sp}_2(\R)$ adapted to the sequence $(A_{n}^{\omega^{(n)}}(E))_{n\in \Z}$. The distribution $\mu_{E}$ is given by: 
\[
\mu_{E}(\Delta)=\nu(\{ \omega^{(0)}=(\omega_{1}^{(0)},\omega_2^{(0)}) \in (\mathrm{supp}\,\nu)^{2} \ |\ A_{0}^{\omega^{(0)}}\!(E)\in \Delta \})
\]
for any Borel subset $\Delta\subset\mathrm{Sp}_2(\R)$. The distribution $\mu_E$ is defined by $A_{0}^{\omega^{(0)}}\!(E)$ alone because the random matrices $A_{n}^{\omega^{(n)}}(E)$ are i.i.d.

We then consider $G_{\mu_{E}}$ the smallest closed subgroup of $\mathrm{Sp}_2(\R)$ generated by the support of $\mu_{E}$. Since $\{0,1\}\subset\supp\nu$, we also have: 
\[
A_{0}^{(0,0)}(E),\ A_{0}^{(1,0)}(E),\ A_{0}^{(0,1)}(E),\
A_{0}^{(1,1)}(E)\in G_{\mu_E}\;.
\]
We want to work with explicit forms of these four transfer matrices. First, we set: 
\begin{equation}\label{defMO}
M_{\omega^{(0)}}=\begin{pmatrix}
\omega_{1}^{(0)} & 1 \\
1 & \omega_2^{(0)}
\end{pmatrix}.
\end{equation}
We begin by writing $A_{0}^{\omega^{(0)}}\!(E)$ as an exponential. To do this we associate to the second order differential system (\ref{system2order}) the following first order differential system: 
\begin{equation}\label{system1order}
Y' =\begin{pmatrix}
0 & I_2 \\
M_{\omega^{(0)}}-E & 0
\end{pmatrix} Y
\end{equation}
with $Y\in \mathcal{M}_{4}(\R)$. If $Y$ is the solution with initial condition $Y(0)=I_{4}$, then $A_{0}^{\omega^{(0)}}\!(E)=Y(1)$. Solving (\ref{system1order}), we get:
\begin{equation}\label{expformABN1}
A_{0}^{\omega^{(0)}}\!(E)= 
\exp\begin{pmatrix}
0 & I_2 \\
M_{\omega^{(0)}}-E & 0
\end{pmatrix}.
\end{equation}
To compute this exponential, we have to compute the successive powers of $M_{\omega^{(0)}}$. To do this, we diagonalize the real symmetric matrix $M_{\omega^{(0)}}$ by an orthogonal matrix $S_{\omega^{(0)}}$:
\[
M_{\omega^{(0)}}=\begin{pmatrix}
\omega_{1}^{(0)} & 1 \\
1 & \omega_2^{(0)}
\end{pmatrix} = S_{\omega^{(0)}} \begin{pmatrix}
\lambda_{1}^{\omega^{(0)}} & 0 \\
0 & \lambda_2^{\omega^{(0)}}
\end{pmatrix}S_{\omega^{(0)}}^{-1},
\]
the eigenvalues $\lambda_2^{\omega^{(0)}}\leq \lambda_{1}^{\omega^{(0)}}$ of $M_{\omega^{(0)}}$ being real. We can compute these eigenvalues and the corresponding matrices $S_{\omega^{(0)}}$ for the different values of $\omega^{(0)} \in \{0,1\}^{2}$. We get: 
\begin{alignat*}{2}
S_{(0,0)}&=\frac{1}{\sqrt{2}}\begin{pmatrix}
1 & 1 \\
1 & -1
\end{pmatrix},&&\lambda_{1}^{(0,0)}=1,\quad\lambda_2^{(0,0)}=-1,\\
S_{(1,1)}&=S_{(0,0)},&&\lambda_{1}^{(1,1)}=2,\quad \lambda_2^{(1,1)}=0,\\
S_{(1,0)}&=
\begin{pmatrix}
\frac{2}{\sqrt{10-2\sqrt{5}}} & \frac{2}{\sqrt{10+2\sqrt{5}}} \\[1mm]
\frac{-1+\sqrt{5}}{\sqrt{10-2\sqrt{5}}} &
\frac{-1-\sqrt{5}}{\sqrt{10+2\sqrt{5}}}
\end{pmatrix},&\quad&\lambda_{1}^{(1,0)}=\frac{1+\sqrt{5}}{2},\
\lambda_2^{(1,0)}=\frac{1-\sqrt{5}}{2},\\
S_{(0,1)}&=
\begin{pmatrix}
\frac{2}{\sqrt{10-2\sqrt{5}}} & \frac{2}{\sqrt{10+2\sqrt{5}}} \\[1mm]
\frac{1-\sqrt{5}}{\sqrt{10-2\sqrt{5}}} &
\frac{1+\sqrt{5}}{\sqrt{10+2\sqrt{5}}}
\end{pmatrix},&&\lambda_{1}^{(0,1)}=\frac{1+\sqrt{5}}{2},\
\lambda_2^{(0,1)}=\frac{1-\sqrt{5}}{2}\;.
\end{alignat*}
We also define the block matrices:  
\[
R_{\omega^{(0)}}=\begin{pmatrix}
S_{\omega^{(0)}} & 0 \\
0 & S_{\omega^{(0)}}
\end{pmatrix}.
\]
Let $E>2$ be larger than all eigenvalues of all $M_{\omega^{(0)}}$. With the abbreviation 
\[
r_l=
r_l(E,\omega^{(0)}):= \sqrt{E-\lambda_l^{\omega^{(0)}}},\quad l=1,2,
\]
the transfer matrices become
\begin{equation} \label{expltransfer}
A_{0}^{\omega^{(0)}}\!(E)=R_{\omega^{(0)}}
\begin{pmatrix}
\cos r_{1}& 0 &\frac{\sin r_1}{r_1} & 0 \\[1mm]
0 &\cos r_{2}& 0 & \frac{\sin r_2}{r_2} \\[1mm]
-r_1\sin r_1 & 0 & \cos r_{1}& 0 \\[1mm]
0 & -r_2\sin r_2 & 0 & \cos r_{2}
\end{pmatrix} R_{\omega^{(0)}}^{-1}\;.
\end{equation}

\section{Proof of Theorem \ref{HAB2thm}}\label{secproof}

We will show in the last part of this section that Theorem \ref{HAB2thm} can be easily deduced from the following proposition: 

\begin{prop}\label{H2zar}
There exists a discrete set $\mathcal{S}_{\mathrm{B}}$ such that for all $E>2$, $E\notin\mathcal{S}_{\mathrm{B}}$, $G_{\mu_E}$ is dense, therefore Zariski-dense, in $\mathrm{Sp}_2(\R)$.
\end{prop}

To prove this proposition, we will follow Theorem \ref{breuillard} for $G=\mathrm{Sp}_2(\R)$.
Let $\mathcal{O}$ be a neighborhood of the identity in $G=\mathrm{Sp}_2(\R)$ as in Theorem~\ref{breuillard}.

\subsection{Elements of $G_{\mu_{E}}$ in $\mathcal{O}$}\label{secdirichlet}

To apply Theorem \ref{breuillard} we need to work with elements in the neighborhood $\mathcal{O}$ of the identity. We will work with the four matrices $A_{0}^{(0,0)}(E)$,  $A_{0}^{(1,0)}(E)$, $A_{0}^{(0,1)}(E)$ and $A_{0}^{(1,1)}(E)$ which are in $G_{\mu_{E}}$. We will prove that by taking a suitable power of each of these matrices we find four matrices in $G_{\mu_{E}}$ which lies in an arbitrary small neighborhood of the identity and thus in $\mathcal{O}$. For this we will use a simultaneous diophantine approximation result.

\begin{thm}[Dirichlet \cite{schmidt}]\label{dirichlet}
Let $\alpha_{1},\ldots,\alpha_{N}$ be real numbers and let $M >1$ be an integer. There exist integers $y,x_{1},\ldots,x_{N}$ in $\Z$ such that $1\leq y\leq M$ and
\[
|\alpha_{i}y-x_{i}|<M^{-\frac{1}{N}}
\]
for $i=1,\ldots,N$. 
\end{thm}

From this theorem we deduce the proposition: 

\begin{prop}\label{diop}
Let $E\in (2,+\infty)$. For all $\omega^{(0)} \in \{0,1\}^{2}$, there exists an integer $m_{\omega}(E)\geq 1$ such that: 
\[
A_{0}^{\omega^{(0)}}\!(E)^{m_{\omega}(E)} \in \mathcal{O}.
\]
\end{prop}
\medskip

\begin{proof}
We fix $\omega^{(0)} \in \{0,1\}^{2}$. Let $M>1$ be an integer. Apply Theorem \ref{dirichlet} 
with $\alpha_{1}=\frac{r_{1}}{2\pi}$ and $\alpha_2=\frac{r_2}{2\pi}$. 
Then there exist $y,x_1,x_2\in \Z$, such that $1\leq y \leq M$ and
\[
\left|\frac{r_{1}}{2\pi}y-x_{1}\right|<M^{-\frac{1}{2}},\quad \left|\frac{r_2}{2\pi}y-x_2\right|<M^{-\frac{1}{2}}, 
\]
which be can be written as: 
\begin{equation}\label{diopapp}
|r_{1}y-2x_{1}\pi|<2\pi M^{-\frac{1}{2}},\quad |r_2y-2x_2\pi|<2\pi M^{-\frac{1}{2}}.
\end{equation}
Let $\theta_i=yr_i-2\pi x_i$, $i=1,2$. Then we have:  
\begin{align*}
A_{0}^{\omega^{(0)}}\!(E)^{y}
&=
R_{\omega^{(0)}}
\begin{pmatrix}
\cos yr_1 & 0 &\frac{\sin yr_1}{r_{1}} & 0\\
0 & \cos yr_2&0 & \frac{\sin yr_2}{r_2}\\\
-r_1\sin yr_1& 0&\cos yr_1 & 0 \\
0 & -r_2\sin yr_2&0 & \cos yr_2
\end{pmatrix}
R_{\omega^{(0)}}^{-1}\\
&= 
R_{\omega^{(0)}}
\begin{pmatrix} 
\cos\theta_1& 0&\frac{\sin\theta_1}{r_{1}} & 0 \\
0 & \cos\theta_2&0&\frac{\sin\theta_2}{r_{2}}\\
-r_{1}\sin\theta_1 & 0&\cos\theta_1 & 0 \ \\
0 & -r_2\sin\theta_2&0&\cos\theta_2
\end{pmatrix}
R_{\omega^{(0)}}^{-1}
\end{align*}
by $2\pi$-periodicity of sinus and cosinus.
Let $\varepsilon >0$. If we choose $M$ large enough, $M^{-\frac{1}{2}}$ will be small enough to get: 
\[
\left\Vert
\begin{pmatrix} 
\cos\theta_1& 0&\frac{\sin\theta_1}{r_{1}} & 0 \\
0 & \cos\theta_2&0&\frac{\sin\theta_2}{r_{2}}\\
-r_{1}\sin\theta_1 & 0&\cos\theta_1 & 0 \ \\
0 & -r_2\sin\theta_2&0&\cos\theta_2
\end{pmatrix}
-I_{4}\,
\right\Vert< \varepsilon.
\]
The matrices $S_{\omega^{(0)}}$ being orthogonal, so are also the matrices $R_{\omega^{(0)}}$. Then conjugating
by $R_{\omega^{(0)}}$ does not change the norm: 
\[
\Vert A_{0}^{\omega^{(0)}}\!(E)^{y}-I_{4}\Vert<\varepsilon.
\]
As $\mathcal{O}$ depends only on the semisimple group $\mathrm{Sp}_2(\R)$, we can choose $\varepsilon$ such that $B(I_{4},\varepsilon)\subset \mathcal{O}$. So if we set $y=m_{\omega}(E)$, we have $1\leq m_{\omega}(E) \leq M$ and:
\[
A_{0}^{\omega^{(0)}}\!(E)^{m_{\omega}(E)} \in \mathcal{O}.
\]
\end{proof}

\begin{rem}
It is important to note that the neighborhood $\mathcal{O}$ does not depend neither on $E$ nor on $\omega^{(0)}$. So the integer $M>1$ also does not depend neither on $E$ nor on $\omega^{(0)}$. It will be important in a next step of the proof to be able to say that even if the integer $m_{\omega}(E)$ depends on $E$ and $\omega^{(0)}$, it belongs always to an interval of integers $\{1,\ldots, M\}$ independent of $E$ and $\omega^{(0)}$. 
\end{rem}

To apply Theorem \ref{breuillard}, we need to show that the logarithms of the matrices $A_{0}^{\omega^{(0)}}\!(E)^{m_{\omega}(E)}$ generate the Lie algebra $\mathfrak{sp}_2(\R)$ of $\mathrm{Sp}_2(\R)$.  A first difficulty is to compute the logarithm of $A_{0}^{\omega^{(0)}}\!(E)^{m_{\omega}(E)}$ which belongs to $\log \mathcal{O}$.

\subsection{Computation of the logarithm of $A_{0}^{\omega}(E)^{m_{\omega}(E)}$}\label{seccalcullog}

We fix $\omega^{(0)} \in \{ 0,1\}^{2}$. We assume $E>2$. Let $\vartheta_i=m_{\omega}(E)r_i$, $i=1,2$.
To compute the logarithm of $A_{0}^{\omega}(E)^{m_{\omega}(E)}$, we start from its expression:
\[
A_{0}^{\omega^{(0)}}\!(E)^{m_{\omega}(E)}=R_{\omega^{(0)}}
 \begin{pmatrix}
\cos\vartheta_1 & 0 & \frac{\sin\vartheta_1}{r_{1}}& 0 \\[1mm]
0 & \cos\vartheta_2 & 0 & \frac{\sin\vartheta_2}{r_2}\\[1mm]
-r_{1} \sin\vartheta_1 & 0 & \cos\vartheta_1 & 0 \\[1mm]
0 & -r_2 \sin\vartheta_2 & 0 & \cos\vartheta_2
\end{pmatrix} R_{\omega^{(0)}}^{-1}.
\]
We can always permute the vectors of the orthonormal basis defined by the columns of $R_{\omega^{(0)}}$. 
So there exists a permutation matrix $P_{\omega^{(0)}}$ (thus orthogonal) such that: 
\begin{align*}
&
A_{0}^{\omega^{(0)}}\!(E)^{m_{\omega}(E)}\\
&\quad
=R_{\omega^{(0)}} P_{\omega^{(0)}} 
 \begin{pmatrix}
\cos\vartheta_1 & \frac{ \sin\vartheta_1}{r_{1}}& 0 & 0 \\[1mm]
-r_{1} \sin\vartheta_1 & \cos\vartheta_1 & 0 & 0 \\[1mm]
0 & 0 & \cos\vartheta_2 & \frac{\sin\vartheta_2}{r_2}\\[1mm]
0 & 0 & -r_2 \sin\vartheta_2 & \cos\vartheta_2
\end{pmatrix} 
P_{\omega^{(0)}}^{-1} R_{\omega^{(0)}}^{-1}.
\end{align*}
Recall that we can choose $m_{\omega}(E)$ such that $A_{0}^{\omega}(E)^{m_{\omega}(E)}$ is 
arbitrarily close to the identity in $\mathrm{Sp}_2(\R)$. Particularly we can assume that: 
\[
\left\Vert A_{0}^{\omega^{(0)}}\!(E)^{m_{\omega}(E)}-I_{4}\right\Vert < 1 \;.
\]
So we can use the power series of the logarithm: 
\begin{equation}\label{logserie}
\log  A_{0}^{\omega^{(0)}}\!(E)^{m_{\omega}(E)}=\sum_{k \geq 1} \frac{(-1)^{k+1}}{k} (A_{0}^{\omega^{(0)}}\!(E)^{m_{\omega}(E)}-I_{4} )^{k}.
\end{equation}
To simplify our computations we will also use the complex forms of sinus and cosinus. We set:
\[
Q_{\omega^{(0)}}= \begin{pmatrix} 
-\frac{\mathrm{i}}{r_{1}} & \frac{\mathrm{i}}{r_{1}} & 0 & 0 \\[1mm]
1 & 1 & 0 & 0 \\[1mm]
0 & 0 & -\frac{\mathrm{i}}{r_2} & \frac{\mathrm{i}}{r_2} \\[1mm]
0 & 0 & 1 & 1
\end{pmatrix}.
\]
Hence: 
\[
Q_{\omega^{(0)}}^{-1}=\frac{1}{2}  \begin{pmatrix} 
\mathrm{i}r_{1} & 1 & 0 & 0 \\
-\mathrm{i}r_{1} & 1 & 0 & 0 \\
0 & 0 & \mathrm{i}r_2 & 1 \\
0 & 0 & -\mathrm{i}r_2 & 1
\end{pmatrix}
\]
Let 
\begin{equation}\label{exp.imomega}
\kappa_l^{\pm}=\ee^{\pm\ii m_{\omega}(E)r_l},\quad l=1,2.
\end{equation} 
Then we have: 
\begin{align*}
&
A_{0}^{\omega^{(0)}}\!(E)^{m_{\omega}(E)}-I_{4}\\ 
&=
R_{\omega^{(0)}}P_{\omega^{(0)}}Q_{\omega^{(0)}}  \begin{pmatrix}
\kappa_1^+-1 & 0 & 0 & 0 \\
0 & \kappa_1^--1 & 0 & 0 \\
0 & 0 & \kappa_2^+-1& 0 \\
0 & 0 & 0 & \kappa_2^--1
\end{pmatrix}
Q_{\omega^{(0)}}^{-1}P_{\omega^{(0)}}^{-1}R_{\omega^{(0)}}^{-1}.
\end{align*}
So by using (\ref{logserie}) we only have to compute: 
\[
\sum_{k=1}^{+\infty}
\frac{(-1)^{k+1}}{k}(\kappa_l^{\pm}-1)^{k}.
\]
Let $\Ln$ be the main determination of the complex logarithm defined on $\C\setminus\R_{-}$. 
We want to write, for $l=1,2$:
\[
\sum_{k=1}^{+\infty}
\frac{(-1)^{k+1}}{k}(\kappa_l^{\pm}-1)^{k}=\Ln\kappa_l^{\pm}.
\]
To do this, we have to assume that $r_{l}=\sqrt{E-\lambda_{l}^{\omega^{(0)}}} \notin\pi+2\pi\Z$. 
So we introduce the discrete set 
\[
\mathcal{S}_{1}=\{E>2\mid E=-\lambda_{l}^{\omega^{(0)}}+\pi+2j\pi\textrm{ for }j\in \Z,\ l=1,2,\ \omega^{(0)} \in \{0,1\}^{2}\}.
\]
If we choose $E>2$, $E\notin\mathcal{S}_{1}$ we can write: 
\begin{align*}
&
\log A_{0}^{\omega^{(0)}}\!(E)^{m_{\omega}(E)}\\ 
&\;=
R_{\omega^{(0)}}P_{\omega^{(0)}}Q_{\omega^{(0)}}  \begin{pmatrix}
\Ln\kappa_1^+& 0 & 0 & 0 \\
0 & \Ln\kappa_1^-& 0 & 0 \\
0 & 0 &\Ln\kappa_2^+ & 0 \\
0 & 0 & 0 & \Ln\kappa_2^-
\end{pmatrix}
Q_{\omega^{(0)}}^{-1}P_{\omega^{(0)}}^{-1}R_{\omega^{(0)}}^{-1}.
\end{align*} 
So we are left with computing $\Ln\kappa_l^{\pm}$. We do this for $l=1$, the computation will be the same for $l=2$. We have: 
\begin{align}                               
\Ln\kappa_1^+ 
&=\mathrm{i}\,\Arg\kappa_1^+
=
\mathrm{i}\,\Arcsin\sin\vartheta_1\nonumber\\
&=
\mathrm{i} \left(m_{\omega}(E)r_{1} - 
 \pi\left\lfloor\frac{m_{\omega}(E)r_{1}}{\pi} + \frac{1}{2} \right\rfloor\right) (-1)^{\left\lfloor\frac{m_{\omega}(E)r_{1}}{\pi} + \frac{1}{2} \right\rfloor} \label{floor}
\end{align}
where $\lfloor\,\cdot\,\rfloor$ in (\ref{floor}) denotes the integer part. We recall that by (\ref{diopapp}), $m_{\omega}(E)r_{1}$ can be chosen arbitrarily close to $2\pi\Z$, i.e.\ we can assume that $\frac{m_{\omega}(E)r_{1}}{\pi}$ is arbitrarily close to an even integer. It suffices to choose $M$ such that $2M^{-\frac{1}{2}}<\frac{1}{2}$ to have
$\left\lfloor\frac{m_{\omega}(E)r_{1}}{\pi} + \frac{1}{2} \right\rfloor$ even and more precisely equal to $2x_{1}$. Thus (\ref{floor}) becomes: 
\begin{equation}\label{floorpos}
\Ln\kappa_1^+ = 
\mathrm{i} \left(m_{\omega}(E)r_{1} - \pi\left\lfloor\frac{m_{\omega}(E)r_{1}}{\pi} + \frac{1}{2}\right\rfloor\right).
\end{equation}
We have the corresponding equation for the conjugate logarithm:  
\begin{equation}\label{floorneg}
\Ln\kappa_1^-
=
\mathrm{i} \left(-m_{\omega}(E)r_{1} - \pi\left\lfloor -\frac{m_{\omega}(E)r_{1}}{\pi} 
+ \frac{1}{2}\right\rfloor\right) (-1)^{\left\lfloor-\frac{m_{\omega}(E)r_{1}}{\pi} + \frac{1}{2} \right\rfloor}.
\end{equation}
We have:
\begin{align}
&
\begin{pmatrix} 
-\frac{\mathrm{i}}{r_{1}} & \frac{\mathrm{i}}{r_{1}} \\[1mm]
1 & 1  \\[1mm]
\end{pmatrix}. \begin{pmatrix}
\Ln\kappa_1^+& 0 \\
0 & \Ln\kappa_1^- \\
\end{pmatrix}\times\frac{1}{2}
\begin{pmatrix} 
\mathrm{i}r_{1} & 1 \\
-\mathrm{i}r_{1} & 1  \\
\end{pmatrix}\nonumber \\ 
&\quad=
\frac{1}{2} \begin{pmatrix}
\Ln\kappa_1^++\Ln\kappa_1^- & -\frac{\mathrm{i}}{r_{1}}\left( \Ln\kappa_1^+ -\Ln\kappa_1^- \right) \\[1mm]
\mathrm{i}r_{1}\left( \Ln\kappa_1^+ -\Ln\kappa_1^- \right) & \Ln\kappa_1^++\Ln\kappa_1^-
\end{pmatrix}\label{RP}
\end{align}
By (\ref{floorpos}) and (\ref{floorneg}) we have: 
\[
\Ln\kappa_1^++\Ln\kappa_1^- 
=-\mathrm{i}\pi \left(\left\lfloor \frac{m_{\omega}(E)r_{1}}{\pi} + \frac{1}{2} \right\rfloor+\left\lfloor -\frac{m_{\omega}(E)r_{1}}{\pi} + \frac{1}{2} \right\rfloor\right)
\]
and, for all $x\in\R$:
\[
\left\lfloor x+\frac{1}{2}\right\rfloor+\left\lfloor \frac{1}{2}-x\right\rfloor=
\begin{cases} 
1 &\text{if }x\in\frac{1}{2}+\Z,\\
0 &\text{otherwise}.
\end{cases}
\]
We can assume that $\frac{m_{\omega}(E)r_{l}}{\pi}$ is arbitrarily close to an even number, 
hence we can assume that for $l=1,2$, $\frac{m_{\omega}(E)r_{l}}{\pi}$ does not belong to $\frac{1}{2}+ \Z$. So we have: 
\begin{equation}\label{lnplus}
\Ln\kappa_1^++\Ln\kappa_1^- = 0
\end{equation}
and:
\begin{align}
&\Ln\kappa_1^+-\Ln\kappa_1^-\nonumber\\
&=2\mathrm{i}m_{\omega}(E)r_{1}
-\mathrm{i}\pi\left( \left\lfloor \frac{m_{\omega}(E)r_{1}}{\pi} - \frac{1}{2} \right\rfloor-\left\lfloor -\frac{m_{\omega}(E)r_{1}}{\pi} + \frac{1}{2} \right\rfloor\right)\nonumber \\ 
&=2\mathrm{i}m_{\omega}(E)r_{1}-2\mathrm{i}\pi\left\lfloor \frac{m_{\omega}(E)r_{1}}{\pi} - \frac{1}{2} \right\rfloor.\label{lnmoins}
\end{align}
Let, for $l=1,2$:
\begin{align}\label{defixi}
x_{l}&=x_{l}(E,\omega):=\frac{1}{2}\left\lfloor \frac{m_{\omega}(E)r_{l}}{\pi} - \frac{1}{2}\right\rfloor.\\
\alpha_l&=-m_{\omega}(E)r_l^{2}+2\pi r_lx_l,\nonumber\\
\beta_l&=m_{\omega}(E)-\frac{2\pi x_l}{r_l}.\nonumber
\end{align}
Putting (\ref{lnplus}) and (\ref{lnmoins}) into (\ref{RP}), 
and doing the same for the block corresponding to $r_2$, we get: 
\begin{align*}
\log A_{0}^{\omega^{(0)}}\!(E)^{m_{\omega}(E)}
&=
R_{\omega^{(0)}}P_{\omega^{(0)}}
 \begin{pmatrix}
0 & \beta_1 & 0 & 0 \\
\alpha_1 & 0 & 0 & 0 \\
0 & 0 & 0 & \beta_2 \\
0 & 0 &\alpha_2& 0 
\end{pmatrix}
P_{\omega^{(0)}}^{-1}R_{\omega^{(0)}}^{-1}\\
&=
R_{\omega^{(0)}} 
 \begin{pmatrix}
0 & 0 & \beta_1 & 0 \\
0 & 0 & 0 & \beta_2 \\
\alpha_1 & 0 & 0 & 0 \\
0 &\alpha_2 & 0 & 0 
\end{pmatrix}
R_{\omega^{(0)}}^{-1} 
\end{align*} 
We set:
\[
LA_{\omega^{(0)}}:=\log A_{0}^{\omega^{(0)}}\!(E)^{m_{\omega}(E)}.
\]
We can summarize the computations we have done in this section. 
For all $E>2$, $E\notin\mathcal{S}_{1}$: 
\begin{equation}\label{exprLA2E}
LA_{\omega^{(0)}}=R_{\omega^{(0)}}  \begin{pmatrix}
0 & 0 & \beta_1 & 0 \\
0 & 0 & 0 & \beta_2 \\
\alpha_1 & 0 & 0 & 0 \\
0 &\alpha_2 & 0 & 0 
\end{pmatrix}R_{\omega^{(0)}}^{-1}.
\end{equation}
We have now to prove that the four matrices $LA_{\omega^{(0)}}$, for $\omega^{(0)}\in \{ 0,1\}^{2}$, 
generate the whole Lie algebra $\mathfrak{sp}_2(\R)$.

\subsection{The Lie algebra $\mathfrak{la}_2(E)$}\label{secLA2E}

For $E\in (2,+\infty)\setminus \mathcal{S}_{1}$, we denote by $\mathfrak{la}_2(E)$ the Lie subalgebra 
of $\mathfrak{sp}_2(\R)$ generated by the $LA_{\omega^{(0)}}$ for $\omega^{(0)}\in \{ 0,1\}^{2}$. We will use the expressions of $\lambda_{i}^{\omega{(0)}}$ and $S_{\omega}$ computed in Section \ref{secmodel}.

\subsubsection{Notations}\label{notationsLA}

We set: 
\begin{align*}
a_{1}&=x_{1}(E,(0,0))=\left\lfloor  \frac{m_{(0,0)}(E)\sqrt{E-1}}{\pi} + \frac{1}{2}\right\rfloor\\ 
a_2&=x_2(E,(0,0))=\left\lfloor  \frac{m_{(0,0)}(E)\sqrt{E+1}}{\pi} + \frac{1}{2}\right\rfloor\\
b_{1}&=x_{1}(E,(1,0))=\left\lfloor  \frac{m_{(1,0)}(E)\sqrt{E-\frac{1+\sqrt{5}}{2}}}{\pi} + \frac{1}{2}\right\rfloor\\
b_2&=x_2(E,(1,0))=\left\lfloor  \frac{m_{(1,0)}(E)\sqrt{E-\frac{1-\sqrt{5}}{2}}}{\pi} + \frac{1}{2}\right\rfloor
\end{align*}
and 
\begin{align*}
c_{1}&=x_{1}(E,(0,1))=\left\lfloor  \frac{m_{(0,1)}(E)\sqrt{E-\frac{1+\sqrt{5}}{2}}}{\pi} + \frac{1}{2}\right\rfloor\\
c_2&=x_2(E,(0,1))=\left\lfloor  \frac{m_{(0,1)}(E)\sqrt{E-\frac{1-\sqrt{5}}{2}}}{\pi} + \frac{1}{2}\right\rfloor\\
d_{1}&=x_{1}(E,(1,1))=\left\lfloor  \frac{m_{(1,1)}(E)\sqrt{E}}{\pi} + \frac{1}{2}\right\rfloor\\ 
d_2&=x_2(E,(1,1))=\left\lfloor  \frac{m_{(1,1)}(E)\sqrt{E-2}}{\pi} + \frac{1}{2}\right\rfloor.
\end{align*}
We denote by $M[i,j]$ the $(i,j)$ entry of a matrix $M$. We also set: 
\begin{align*}
r_{1}^{00}&=\sqrt{E-1},\qquad\qquad\qquad\quad\;\; r_2^{00}=\sqrt{E+1},\\
r_{1}^{11}&=\sqrt{E-2},\qquad\qquad\qquad\quad\;\;  r_2^{11}=\sqrt{E},\\
r_{1}^{10}&=r_{1}^{01}=\sqrt{E-\frac{1+\sqrt{5}}{2}},\qquad r_2^{10}=r_2^{01}:=\sqrt{E-\frac{1-\sqrt{5}}{2}},
\end{align*}
and finally we set: 
\begin{align*}
D_{1}(E) &= \sqrt{E-1}\sqrt{E+1}\sqrt{E-\frac{1+\sqrt{5}}{2}}\sqrt{E-\frac{1-\sqrt{5}}{2}},\\
D_2(E) &= \sqrt{E}\sqrt{E-2}\sqrt{E-\frac{1+\sqrt{5}}{2}}\sqrt{E-\frac{1-\sqrt{5}}{2}}.
\end{align*}
To prove that $\mathfrak{la}_2(E)=\mathfrak{sp}_2(\R)$, we will find a family of $10$ matrices linearly independent in $\mathfrak{la}_2(E)$. First we will consider the subspace generated by the Lie brackets $[LA_{\omega^{(0)}},LA_{\tilde{\omega}^{(0)}}]$. 

\subsubsection{The subspace $V_{1}$ generated by 
the $[LA_{\omega^{(0)}},LA_{\tilde{\omega}^{(0)}}]$}\label{bracketLA}

A direct computation shows that each Lie bracket $[LA_{\omega^{(0)}},LA_{\tilde{\omega}^{(0)}}]$ is of the form
\begin{equation}\label{formeV1}
\begin{pmatrix}
A & 0 \\
0 & -{^t}A
\end{pmatrix}
\end{equation}
for some $A\in \mathcal{M}_2(\R)$. Let $V_1$ be the $4$-dimensional subspace of $\mathfrak{sp}_2(\R)$ of matrices of the form (\ref{formeV1}). We will show that outside a discrete set of energies $E$, the four Lie brackets 
\begin{alignat*}{2}
\Upsilon_1&=[LA_{(1,0)},LA_{(0,0)}],&&\Upsilon_2=[LA_{(1,0)},LA_{(1,1)}],\\
\Upsilon_3&=[LA_{(0,1)},LA_{(0,0)}],&\qquad&\Upsilon_4=[LA_{(0,1)},LA_{(0,0)}] 
\end{alignat*}
generate $V_{1}$.
\paragraph*{\emph{Expression of} $\Upsilon_1=[LA_{(1,0)},LA_{(0,0)}]$.}
We give the expressions of the entries. By (\ref{formeV1}) it suffices to give 
the entries corresponding to the first diagonal $2\times 2$ block. 
\begin{align*}
\Upsilon_1[1,1]
&=
-\frac{1}{4\sqrt{5}D_{1}(E)} \left[ (-\pi (a_{1} r_2^{00} + a_2 r_{1}^{00})+2m_{00}r_{1}^{00}r_2^{00})\right.\\
&\quad\left. 
(\pi b_{1} (1+\sqrt{5}) r_2^{10}-\pi b_2 (1-\sqrt{5}) r_{1}^{10} -2\sqrt{5}m_{10}  r_{1}^{10} r_2^{10}) \right]\\
\Upsilon_1[1,2]  
&= 
\frac{\pi^{2} E}{2\sqrt{5}D_{1}(E)} \left[  (b_{1}r_2^{10}- b_2 r_{1}^{10}) (a_{1}r_2^{00}-a_2r_{1}^{00}) \right]
\end{align*}
\begin{align*}
\Upsilon_1[2,1]
&=
-\frac{\pi}{4\sqrt{5}D_{1}(E)}\times
\left[ \pi (a_2r_{1}^{00}-5a_{1}r_2^{00}+4m_{00}r_{1}^{00}r_2^{00})(b_{1}r_2^{10}+ b_2r_{1}^{10})\right.\\ 
&\quad\left. 
+(a_{1}r_2^{00}-a_2r_{1}^{00}) (2\sqrt{5} m_{10} r_{1}^{10}r_2^{10} + 2\pi E (b_{1}r_2^{10}- b_2r_{1}^{10}) )\right]
\end{align*}
\begin{align*}
\Upsilon_1[2,2]  
&= 
\frac{\pi^{2}}{2\sqrt{5}D_{1}(E)} \left[  (b_{1}r_2^{10}- b_2 r_{1}^{10}) (a_{1}r_2^{00}-a_2r_{1}^{00}) \right].
\end{align*}
\paragraph*{\emph{Expression of} $\Upsilon_2=[LA_{(0,1)},LA_{(0,0)}]$.}
We have: 
\begin{align*}
\Upsilon_2[1,1] 
&=
-\frac{1}{20D_{1}(E)}\times\\
&
\quad\left[ (10\sqrt{5}\pi m_{00}r_{1}^{00}r_2^{00} - \sqrt{5}\pi^{2} (a_2r_{1}^{00}+3a_{1}r_2^{00}) )(c_{1}r_2^{10}-c_2r_{1}^{10})\right.\\[1mm]
&\quad\left.
+  5(\pi^{2}(a_{1}r_2^{00}
-3a_2r_{1}^{00})+2\pi m_{00} r_{1}^{00}r_2^{00})  (c_{1}r_2^{10}+c_2r_{1}^{10})\right.\\[1mm]
&\quad\left.
-10(\pi m_{01}(a_{1}r_2^{00}-3a_2r_{1}^{00})+ 2m_{00}m_{01} r_{1}^{00}r_2^{00})r_{1}^{10}r_2^{10}\right]
\end{align*}
\begin{align*}
\Upsilon_2[1,2]
&=
-\frac{1}{2\sqrt{5}D_{1}(E)} \left[ ( \pi^{2}(a_{1}r_2^{00}-3a_2r_{1}^{00})+\pi^{2}E(a_{1}r_2^{00}-a_2r_{1}^{00}) \right.\\[1mm]
&\quad\left.
+(2+2\sqrt{5})\pi m_{00}r_{1}^{00}r_2^{00} ) (c_{1}r_2^{10}-c_2r_{1}^{10})\right.\\[1mm]
&\quad\left.
-\sqrt{5}\pi^{2} (a_{1}r_2^{00}+a_2r_{1}^{00})(c_{1}r_2^{10}+c_2r_{1}^{10})\right.\\[1mm]
&\quad\left.
+ 2\sqrt{5}(\pi m_{01} (a_{1}r_2^{00}
+a_2r_{1}^{00}) -2m_{00}m_{01}r_{1}^{00}r_2^{00} ) r_{1}^{10}r_2^{10}\right]
\end{align*}
\begin{align*}
\Upsilon_2[2,1]
&=
-\frac{1}{20D_{1}(E)}\times\\
&
\quad\left[ (5\pi^{2}(a_{1}r_2^{00} + 3a_2r_{1}^{00})-20\pi m_{00}r_{1}^{00}r_2^{00} ) (c_{1}r_2^{10}+c_2r_{1}^{10}) \right.\\[1mm]
&\quad\left.
+\sqrt{5}\pi^{2}(2E-5)(a_{1}r_2^{00}-a_2r_{1}^{00}) (c_{1}r_2^{10}-c_2r_{1}^{10})  \right.\\[1mm]
&\quad\left.
-10(\pi m_{01} (a_{1}r_2^{00}+3a_2r_{1}^{00}) -4m_{00}m_{01}r_{1}^{00}r_2^{00} )r_{1}^{10}r_2^{10} \right]
\end{align*}
\begin{align*}
\Upsilon_2[2,2]
&= 
-\frac{\pi}{10D_{1}(E)} \left[ 5\pi(a_{1}r_2^{00} - a_2r_{1}^{00}) (c_{1}r_2^{10}+c_2r_{1}^{10}) \right.\\[1mm]
&\quad\left.
+2\sqrt{5}(\pi (a_{1}r_2^{00}+a_2r_{1}^{00})+2m_{00}r_{1}^{00}r_2^{00} ) (c_{1}r_2^{10}-c_2r_{1}^{10})\right.\\[1mm]
&\quad\left. 
+10m_{01} (a_{1}r_2^{00}+a_2r_{1}^{00}) r_{1}^{10}r_2^{10} \right].
\end{align*} 
\paragraph*{\emph{Expression of} $\Upsilon_3=[LA_{(1,0)},LA_{(1,1)}]$.}
We have: 
\begin{align*}
\Upsilon_3[1,1]
&=
-\frac{\pi}{10D_2(E)}\times\\
&\quad
\left[ 2\sqrt{5}(2m_{11}r_{1}^{11}r_2^{11}-\pi(d_{1}r_2^{11}+d_2r_{1}^{11}))(b_{1}r_2^{10}-b_2r_{1}^{10}) \right.\\[1mm]
&\quad\left.
+5\pi(d_2r_{1}^{11}-d_{1}r_2^{11})(b_{1}r_2^{10}+b_2r_{1}^{10})\right.\\[1mm]
&\quad\left.
+10m_{10}(d_{1}r_2^{11}-d_2r_{1}^{11})r_{1}^{10}r_2^{10}  \right] 
\end{align*}
\begin{align*}
\Upsilon_3[1,2]
&=
\frac{1}{20D_2(E)} \left[ \sqrt{5}\pi^{2} (d_{1}r_2^{11}-d_2r_{1}^{11})(2E-3)(b_{1}r_2^{10}-b_2r_{1}^{10}) \right.\\[1mm]
&\quad\left.
+(5\pi^{2}(d_{1}r_2^{11}+3d_2r_{1}^{11})-20\pi m_{11}r_{1}^{11}r_2^{11}) (b_{1}r_2^{10}+b_2r_{1}^{10}) \right.\\[1mm]
&\quad\left.
+(40m_{11}m_{10}r_{1}^{11}r_2^{11}-10\pi m_{10}(d_{1}r_2^{11}+3d_2r_{1}^{11}) )r_{1}^{10}r_2^{10} \right] 
\end{align*}
\begin{align*}
\Upsilon_3[2,1]
&=
-\frac{1}{10D_2(E)} \left[ (2\pi^{2}\sqrt{5}(2d_2r_{1}^{11}-d_{1}r_2^{11})\right.\\[1mm]
&\quad\left.
+\pi^{2}\sqrt{5}E(d_{1}r_2^{11}-d_2r_{1}^{11}) -2\pi \sqrt{5}m_{11}r_{1}^{11}r_2^{11})(b_{1}r_2^{10}-b_2r_{1}^{10})\right.\\[1mm]
&\quad\left.
+(10\pi m_{11} r_{1}^{11}r_2^{11}-5\pi^{2}(d_{1}r_2^{11}+d_2r_{1}^{11}))  \right.\\[1mm]
&\quad\left.
(b_{1}r_2^{10}+b_2r_{1}^{10}) +(10\pi m_{10}(d_{1}r_2^{11}+d_2r_{1}^{11})\right.\\[1mm]
&\quad\left.
-20m_{11}m_{00}r_{1}^{11}r_2^{11}) r_{1}^{10}r_2^{10} \right]
\end{align*}
\begin{align*}
\Upsilon_3[2,2]
&=
-\frac{1}{20D_2(E)}\times\\
&\quad
\left[ (10\pi \sqrt{5} m_{11} r_{1}^{11}r_2^{11} 
-\pi^{2} \sqrt{5} (3d_{1}r_2^{11}+7d_2r_{1}^{11}))(b_{1}r_2^{10}-b_2r_{1}^{10}) \right.\\[1mm]
&\quad\left.
+(5\pi^{2} (3d_2r_{1}^{11}-d_{1}r_2^{11})
-10\pi m_{11} r_{1}^{11}r_2^{11}) (b_{1}r_2^{10}+b_2r_{1}^{10})\right.\\[1mm]
&\quad\left. 
+(10\pi m_{10}(d_{1}r_2^{11}-3d_2r_{1}^{11})+20m_{11}m_{00}r_{1}^{11}r_2^{11}) r_{1}^{10}r_2^{10}\right].
\end{align*} 
\paragraph*{\emph{Expression of} $\Upsilon_4=[LA_{(0,1)},LA_{(1,1)}]$.}
We have: 
\begin{align*}
\Upsilon_4[1,1]
&=
\frac{\pi^{2}}{2\sqrt{5}D_2(E)} \left[ (d_{1}r_2^{11}-d_2r_{1}^{11}) (c_{1}r_2^{10}-c_2r_{1}^{10}) \right]\\
\Upsilon_4[1,2]
&=
 \frac{\pi}{4\sqrt{5}D_2(E)}\times\\[1mm]
&\quad
 \left[ (\pi (d_{1}r_2^{11}+3d_2r_{1}^{11}) +2\pi E(d_{1}r_2^{11}-d_2r_{1}^{11})-4m_{11}r_{1}^{11} r_2^{11})  \right.\\[1mm]
&\quad
\left.
(c_{1}r_2^{10}-c_2r_{1}^{10})  +\sqrt{5}\pi (d_2r_{1}^{11}-d_{1}r_2^{11}) (c_{1}r_2^{10}+c_2r_{1}^{10})  \right.\\[1mm]
 &\quad
\left.
+ 2\sqrt{5} m_{01} (d_{1}r_2^{11}-d_2r_{1}^{11})r_{1}^{10}r_2^{10} \right]
\end{align*}
\begin{align*}
\Upsilon_4[2,1]
&=
-\frac{\pi^{2}}{2\sqrt{5}D_2(E)} \left[ (E-1) (d_{1}r_2^{11}-d_2r_{1}^{11}) (c_{1}r_2^{10}-c_2r_{1}^{10})\right]\\ 
\Upsilon_4[2,2]
&=
-\frac{1}{4\sqrt{5}D_2(E)}\times\\[1mm]
&\quad
\left[ (2m_{11}r_{1}^{11}r_2^{11} -\pi (d_{1}r_2^{11}+d_2r_{1}^{11})) (2\sqrt{5}m_{01}r_{1}^{10}r_2^{10}  \right.\\[1mm]
&\quad
\left.
+\pi (c_{1}r_2^{10}-c_2r_{1}^{10}) -\sqrt{5}\pi (c_{1}r_2^{10}+c_2r_{1}^{10}))  \right].
\end{align*} 
To prove that $\mathfrak{la}_2(E)=\mathfrak{sp}_2(\R)$, we will build a family of $10$ matrices linearly independent in $\mathfrak{la}_2(E)$. First we will consider the subspace generated by the Lie brackets $[LA_{\omega^{(0)}},LA_{\tilde{\omega}^{(0)}}]$. 

We can then consider the determinant of these entries: 
\begin{align}\label{detbracket1}
&\det
\begin{pmatrix}
\Upsilon_1[1,1] &\Upsilon_2[1,1] & \Upsilon_3[1,1] &  \Upsilon_4[1,1] \\
\Upsilon_1[1,2] &\Upsilon_2[1,2] & \Upsilon_3[1,2] &  \Upsilon_4[1,2]\\
\Upsilon_1[2,1] &\Upsilon_2[2,1] & \Upsilon_3[2,1] & \Upsilon_4[2,1]\\
\Upsilon_1[2,2] &\Upsilon_2[2,2] & \Upsilon_3[2,2] & \Upsilon_4[2,2]
\end{pmatrix}\\
&=f_1(E)
=\tilde{f}_{1}(a_{1},a_2,b_{1},b_2,c_{1},c_2,d_{1},d_2,m_{00},m_{01},m_{10},m_{11} ,E)\nonumber
\end{align}
where $\tilde{f}_{1}(X_{1},\ldots,X_{12},Y)$ is a polynomial function in $X_{1},\ldots,X_{12}$, analytic in $Y$. Indeed, the determinant (\ref{detbracket1}) is a rational function in the $r_{i}^{jk}$ which are analytic functions in $E$ not vanishing on the interval $(2,+\infty)$. 

Note that all coefficients $a_{1},\ldots,d_2,m_{00},\ldots,m_{11}$ depend also on $E$ and are not analytic in $E$. Hence $f_{1}$ is a priori not analytic in $E$. We will now explain how to avoid this difficulty.

We recall that for all $E$ and $\omega$, $1\leq m_{\omega}(E) \leq M$ with $M$ independent of $E$ and $\omega$. Thus $m_{\omega}(E)$ only take a finite number of values in the set $\{1,\ldots, M\}$.

Then we consider the sequence of intervals $I_2=(2,3]$, $I_{3}=[3,4]$, and for all $k\geq 3$, $I_{k}=[k,k+1]$. These intervals cover $(2,+\infty)$. We fix $k\geq 2$ and we assume that $E\in I_{k}$. Then the integers 
\[
x_{i}^{\omega}(E)=\left\lfloor  \frac{m_{\omega} (E)\sqrt{E-\lambda_{i}^{\omega}}}{\pi} + \frac{1}{2}\right\rfloor
\]
are bounded by a constant depending only on $M$ and $I_{k}$. Indeed, the eigenvalues $\lambda_{i}^{\omega}$ are all in the fixed interval $[-2,2]$, $m_{\omega}(E)$ take its values in $\{1,\ldots, M\}$ and $E\in I_{k}$. So the integers $x_{i}^{\omega}(E)$ take only a finite number of values in a set $\{0,\ldots ,N_{k} \}$. 

To study the zeros of the function $f_{1}$ on $I_{k}$, we have only to study the zeros of a finite number of analytic functions:
\[
\tilde{f}_{1,p,l}\colon E \mapsto \tilde{f}_{1}(p_{1},\ldots ,p_{8},l_{1},\ldots,l_{4},E)
\]
for $p_{i} \in \{0,\ldots, N_{k}\}$ and $l_{j} \in \{1,\ldots,M \}$. We have to show that the functions $\tilde{f}_{1,p,l}$ do not vanish identically on $I_{k}$. In fact, the only bad case is when all the $x_{i}^{\omega}$ are zero. Indeed, $\tilde{f}_{1}(0,\ldots,0,X_{9},\ldots,X_{12},Y)$ is identically zero. But if we look at the values of $x_{i}^{\omega}$ for $E>2$ and $m_{\omega}(E) \geq 1$, we get that $a_2\geq 1$. We can compute the term of the determinant (\ref{detbracket1}) involving only $a_2$. We get:
\[
\frac{m_{10}^{2}m_{01}^{2}m_{11}^{2}\pi^{2}a_2^{2}}{E+1} \geq \frac{\pi^{2}}{E+1} >0.
\]
By observing all entries of (\ref{detbracket1}), this term is the only one involving $E$ only by this power of $E+1=(r_2^{00})^{2}$ and no other power of the $r_{j}^{kl}$. So this term cannot be cancelled uniformly in $E$ by another term of the development of the determinant (\ref{detbracket1}), whatever values taken by the integers $a_{1},b_{1},\ldots,d_2$ and $m_{00},\ldots, m_{11}$. So the only case where $\tilde{f}_{1,p,l}$ could identically vanish does not happen. We set: 
\begin{align*}
J_1
&=
\{(a_1,b_1,\dots,d_2,m_{00},\dots,m_{11})\mid 0\leq a_1,c_1,\dots,d_2\leq N_k,\\
&\qquad\qquad\qquad\qquad\qquad\qquad\quad\;\; 1\leq b_1\leq N_k,\ 1\leq m_{ij}\leq M\}.
\end{align*}
Then, as $(a_{1},\ldots, m_{11})\in J_{1}$ the set of zeros of $f_{1}$ in $I_{k}$ is included in the following finite union of discrete sets: 
\[
\{E \in I_{k} \ |\ f_{1}(E)=0\} \subset \bigcup_{(p,l)\in J_{1}} \{ E\in I_{k} \ |\ \tilde{f}_{1,p,l}(E)=0 \}
\]
Thus this set is also discrete in $I_{k}$. We finally get that:
\[
\{E\in(2,+\infty[\ |\ f_{1}(E)=0 \} = \bigcup_{k\geq 2} \{E \in I_{k} \ |\ f_{1}(E)=0\}
\]
is discrete in $(2,+\infty)$. 
We set: 
\[
\mathcal{S}_2=\{ E>2\mid\ f_{1}(E)=0 \}.
\]

Let $E>2$, $E\notin\mathcal{S}_{1}\cup\mathcal{S}_2$. As the determinant (\ref{detbracket1}) is not zero, it follows that the four matrices $\Upsilon_1,\dots,\Upsilon_4$ are linearly independent in the subspace $V_{1}\subset \mathfrak{sp}_2(\R)$ of dimension $4$. Thus, they generate $V_{1}$. We deduce that: 
\begin{equation}
\textrm{for all } E\in (2,+\infty)\setminus (\mathcal{S}_{1}\cup \mathcal{S}_2),\ V_{1} \subset \mathfrak{la}_2(E)
\end{equation}

We now have to find another family of six matrices linearly independent in a complement of $V_{1}$ in $\mathfrak{sp}_2(\R)$.

\subsubsection{The orthogonal $V_2$ of $V_{1}$ in $\mathfrak{sp}_2(\R)$}\label{diffLA}

We begin by giving the expressions of the three matrices 
\begin{equation}\label{three.brackets}
LA_{(1,0)}-LA_{(0,0)},\ LA_{(1,0)}-LA_{(1,1)},\ LA_{(0,1)}-LA_{(0,0)}. 
\end{equation}
Looking at the form of $LA_{\omega^{(0)}}$ given by (\ref{exprLA2E}) we already know that all these differences are of the form:
\begin{equation}\label{formV2}
\begin{pmatrix}
0 & 0 & e & g \\
0 & 0 & g & f \\
a & c & 0 & 0 \\
c & b & 0 & 0
\end{pmatrix}
\end{equation}
for $(a,b,c,e,f,g)\in\R^{6}$
Let $V_2\subset\mathfrak{sp}_2(\R)$ be the $6$-dimensional subspace of matrices of the form (\ref{formV2}). 
We have $\mathfrak{sp}_2(\R)=V_{1} \oplus V_2$. By (\ref{formV2}) it suffices to compute the 
$[3,1]$, $[3,2]$, $[4,2]$, $[1,3]$,$[1,4]$ and $[2,4]$ entries of the three matrices (\ref{three.brackets}).

\paragraph*{\emph{Expression of} $\Theta_1=LA_{(1,0)}-LA_{(0,0)}$.} 
We have: 
\begin{align*}
\Theta_1[3,1]
&=
m_{10}(1-E)+m_{00}E-\frac{\pi}{2}(a_{1}r_2^{00}+a_2r_{1}^{00})\\
&\quad
+ \frac{\pi}{2\sqrt{5}} (b_{1}r_2^{10}-b_2r_{1}^{10})+ \frac{\pi}{2}(b_{1}r_2^{10}+b_2r_{1}^{10})\\
\Theta_1[3,2]&=
m_{10}-m_{00}+\frac{\pi}{2}(a_2r_{1}^{00}-a_{1}r_2^{00}) + \frac{\pi}{\sqrt{5}} (b_{1}r_2^{10}-b_2r_{1}^{10})\\
\Theta_1[4,2]
&=
(m_{00}-m_{10})E-\frac{\pi}{2}(a_{1}r_2^{00}+a_2r_{1}^{00})\\
&\quad
- \frac{\pi}{2\sqrt{5}} (b_{1}r_2^{10}-b_2r_{1}^{10})+ \frac{\pi}{2}(b_{1}r_2^{10}+b_2r_{1}^{10})\\
\Theta_1[1,3]
&=
m_{10}-m_{00}+\frac{\pi}{2}\left(\frac{a_{1}}{r_2^{00}} -\frac{a_2}{r_{1}^{00}}\right)\\
&\quad
+ \frac{\pi}{2\sqrt{5}} \left(\frac{b_2}{r_{1}^{10}} -\frac{b_{1}}{r_2^{10}} \right)-\frac{\pi}{2} \left(\frac{b_{1}}{r_2^{10}}+ \frac{b_2}{r_{1}^{10}} \right)\\
\Theta_1[1,4]
&=
\frac{\pi}{2}\left(\frac{a_{1}}{r_2^{00}} -\frac{a_2}{r_{1}^{00}}\right) + \frac{\pi}{\sqrt{5}} \left(\frac{b_2}{r_{1}^{10}} -\frac{b_{1}}{r_2^{10}} \right)\\
\Theta_1[2,4]
&=
m_{10}-m_{00}+\frac{\pi}{2}\left(\frac{a_{1}}{r_2^{00}} +\frac{a_2}{r_{1}^{00}}\right)\\
&\quad
+ \frac{\pi}{2\sqrt{5}} \left(\frac{b_{1}}{r_2^{10}}- \frac{b_2}{r_{1}^{10}} \right)-\frac{\pi}{2} \left(\frac{b_{1}}{r_2^{10}}+ \frac{b_2}{r_{1}^{10}} \right)
\end{align*}

\paragraph*{\emph{Expression of} $\Theta_2=LA_{(1,0)}-LA_{(1,1)}$.} 
We have: 
\begin{align*}
\Theta_2[3,1]
&=
m_{10}+(m_{11}-m_{10})E-\frac{\pi}{2}(d_{1}r_2^{11}+d_2r_{1}^{11})\\
&\quad
+ \frac{\pi}{2\sqrt{5}} (b_{1}r_2^{10}-b_2r_{1}^{10})+ \frac{\pi}{2}(b_{1}r_2^{10}+b_2r_{1}^{10})\\
\Theta_2[3,2]
&=
m_{10}+m_{11}+\frac{\pi}{2}(d_2r_{1}^{11}-d_{1}r_2^{11}) + \frac{\pi}{\sqrt{5}} (b_{1}r_2^{10}-b_2r_{1}^{10})\\
\Theta_2[4,2]
&=
(m_{11}-m_{10})E-\frac{\pi}{2}(d_{1}r_2^{11}+d_2r_{1}^{11})\\
&\quad
- \frac{\pi}{2\sqrt{5}} (b_{1}r_2^{10}-b_2r_{1}^{10})+ \frac{\pi}{2}(b_{1}r_2^{10}+b_2r_{1}^{10})\\
\Theta_2[1,3]
&=
m_{10}-m_{11}+\frac{\pi}{2}\left(\frac{d_{1}}{r_2^{00}} +\frac{d_2}{r_{1}^{00}}\right)\\
&\quad
- \frac{\pi}{2\sqrt{5}} \left(\frac{b_{1}}{r_2^{10}}- \frac{b_2}{r_{1}^{10}} \right)-\frac{\pi}{2} \left(\frac{b_{1}}{r_2^{10}}+ \frac{b_2}{r_{1}^{10}} \right)\\
\Theta_2[1,4]
&=
\frac{\pi}{2}\left(\frac{d_{1}}{r_2^{00}} -\frac{d_2}{r_{1}^{00}}\right) - \frac{\pi}{\sqrt{5}} \left(\frac{b_{1}}{r_2^{10}}- \frac{b_2}{r_{1}^{10}} \right)\\
\Theta_2[2,4]
&=
m_{10}-m_{11}+\frac{\pi}{2}\left(\frac{d_{1}}{r_2^{00}} +\frac{d_2}{r_{1}^{00}}\right)\\
&\quad
+ \frac{\pi}{2\sqrt{5}} \left(\frac{b_{1}}{r_2^{10}}- \frac{b_2}{r_{1}^{10}} \right)-\frac{\pi}{2} \left(\frac{b_{1}}{r_2^{10}}+ \frac{b_2}{r_{1}^{10}} \right)
\end{align*}

\paragraph*{\emph{Expression of} $\Theta_3=LA_{(0,1)}-LA_{(0,0)}$.} 
We have: 
\begin{align*}
\Theta_3[3,1]
&=
m_{01}+(m_{00}-m_{01})E-\frac{\pi}{2}(a_{1}r_2^{00}+a_2r_{1}^{00})\\
&\quad
+ \frac{\pi}{2\sqrt{5}} (c_{1}r_2^{10}-c_2r_{1}^{10})+ \frac{\pi}{2}(c_{1}r_2^{10}+c_2r_{1}^{10})\\
\Theta_3[3,2]
&=
-(m_{00}+m_{01})+\frac{\pi}{2}(a_2r_{1}^{00}-a_{1}r_2^{00})\\
&\quad
+ \frac{\pi}{\sqrt{5}} (c_2r_{1}^{10}-c_{1}r_2^{10})\\
\Theta_3[4,2]
&=
(m_{00}-m_{01})E-\frac{\pi}{2}(a_{1}r_2^{00}+a_2r_{1}^{00})\\
&\quad
+ \frac{\pi}{2\sqrt{5}} (c_{1}r_2^{10}-c_2r_{1}^{10})- \frac{\pi}{2}(c_{1}r_2^{10}+c_2r_{1}^{10})\\
\Theta_3[1,3]
&=
m_{01}-m_{00}+\frac{\pi}{2}\left(\frac{a_{1}}{r_2^{00}}+\frac{a_2}{r_{1}^{00}}\right)\\
&\quad
+ \frac{\pi}{2\sqrt{5}} \left(\frac{c_2}{r_{1}^{10}}-\frac{c_{1}}{r_2^{10}} \right)- \frac{\pi}{2}\left( \frac{c_{1}}{r_2^{10}}+\frac{c_2}{r_{1}^{10}} \right)\\
\Theta_3[1,4]
&=
\frac{\pi}{2}\left(\frac{a_{1}}{r_2^{00}}-\frac{a_2}{r_{1}^{00}}\right)
- \frac{\pi}{\sqrt{5}} \left(\frac{c_2}{r_{1}^{10}}-\frac{c_{1}}{r_2^{10}} \right)\\
\Theta_3[2,4]
&=
m_{01}-m_{00}+\frac{\pi}{2}\left(\frac{a_{1}}{r_2^{00}}+\frac{a_2}{r_{1}^{00}}\right)\\
&\quad
- \frac{\pi}{2\sqrt{5}} \left(\frac{c_2}{r_{1}^{10}}-\frac{c_{1}}{r_2^{10}} \right)- \frac{\pi}{2}\left( \frac{c_{1}}{r_2^{10}}+\frac{c_2}{r_{1}^{10}} \right)
\end{align*}

Now we assume that $E\in (2,+\infty) \setminus (\mathcal{S}_{1}\cup \mathcal{S}_2)$. Then $V_{1} \subset \mathfrak{la}_2(E)$ and in particular the following matrices are in $\mathfrak{la}_2(E)$: 
\[
Z_{1}= \begin{pmatrix}
1 & 0 & 0 & 0 \\
0 & 0 & 0 & 0 \\
0 & 0 & -1 & 0 \\
0 & 0 & 0 & 0 
\end{pmatrix},\ Z_2= \begin{pmatrix}
0 & 0 & 0 & 0 \\
0 & 1 & 0 & 0 \\
0 & 0 & 0 & 0 \\
0 & 0 & 0 & -1 
\end{pmatrix},\ Z_{3}= \begin{pmatrix}
0 & 1 & 0 & 0 \\
0 & 0 & 0 & 0 \\
0 & 0 & 0 & 0 \\
0 & 0 & -1 & 0
\end{pmatrix}.
\]
So we can consider the three matrices of $\mathfrak{la}_2(E)$:
\[
[LA_{(1,0)}-LA_{(0,0)},Z_{1}],\quad 
[LA_{(1,0)}-LA_{(1,1)},Z_2],\quad
[LA_{(0,1)}-LA_{(0,0)},Z_{3}].
\] 
We can check that in general the Lie bracket of an element of $V_{1}$ and an element of $V_2$ is still in $V_2$. So, to write this three matrices we will only have to give explicitly six of their entries.

\paragraph*{\emph{Expression of} $\Theta_4=[LA_{(1,0)}-LA_{(0,0)},Z_{1}]$.} 
We have:
\begin{align*}
\Theta_4[3,1]
&=
2m_{10}+2(m_{00}-m_{10})E-\pi (a_{1}r_2^{00}+a_2r_{1}^{00})\\
&\quad
+\pi(b_{1}r_2^{10}+b_2r_{1}^{10})+\frac{\pi}{\sqrt{5}}(b_{1}r_2^{10}-b_2r_{1}^{10})\\
\Theta_4[3,2]
&=
m_{10}-m_{00}+\pi (a_2r_{1}^{00}-a_{1}r_2^{00})+\frac{\pi}{\sqrt{5}}(b_{1}r_2^{10}-b_2r_{1}^{10})\\
\Theta_4[4,2]
&=
0\\
\Theta_4[1,3]
&=
2(m_{00}-m_{10})-\pi \left(\frac{a_{1}}{r_2^{00}}+\frac{a_2}{r_{1}^{00}}\right)\\
&\quad
+\pi \left( \frac{b_{1}}{r_2^{10}}+\frac{b_2}{r_{1}^{10}}\right)+\frac{\pi}{\sqrt{5}}\left( \frac{b_{1}}{r_2^{10}}-\frac{b_2}{r_{1}^{10}}\right)\\
\Theta_4[1,4]
&=
\frac{\pi}{2}\left(\frac{a_2}{r_{1}^{00}}- \frac{a_{1}}{r_2^{00}} \right)+\frac{\pi}{\sqrt{5}}\left( \frac{b_{1}}{r_2^{10}}-\frac{b_2}{r_{1}^{10}}\right)\\
\Theta_4[2,4]
&=
0.
\end{align*}

\paragraph*{\emph{Expression of} $\Theta_5=[LA_{(1,0)}-LA_{(1,1)},Z_2]$.}
We have:
\begin{align*}
\Theta_5[3,1]
&=
0\\
\Theta_5[3,2]
&=
m_{10}+m_{11}+\frac{\pi}{2}(d_2r_{1}^{11}-d_{1}r_2^{11}) +\frac{\pi}{\sqrt{5}}(b_{1}r_2^{10}-b_2r_{1}^{10})\\
\Theta_5[4,2]
&=
2(m_{11}-m_{10})E-2m_{11}-\pi (d_{1}r_2^{11}+d_2r_{1}^{11})\\
&\quad
+\pi (b_{1}r_2^{10}+b_2r_{1}^{10})-\frac{\pi}{\sqrt{5}}(b_{1}r_2^{10}-b_2r_{1}^{10})\\
\Theta_5[1,3]
&=
0\\
\Theta_5[1,4]
&=
\frac{\pi}{2} \left(\frac{d_2}{r_{1}^{11}}- \frac{d_{1}}{r_2^{11}} \right)+\frac{\pi}{\sqrt{5}}\left( \frac{b_{1}}{r_2^{10}}-\frac{b_2}{r_{1}^{10}}\right)\\
\Theta_5[2,4]
&=
2(m_{11}-m_{10})-\pi \left( \frac{d_{1}}{r_2^{11}}+\frac{d_2}{r_{1}^{11}}\right)\\
&\quad
+\pi \left( \frac{b_{1}}{r_2^{10}}+\frac{b_2}{r_{1}^{10}}\right)  -\frac{\pi}{\sqrt{5}}\left( \frac{b_{1}}{r_2^{10}}-\frac{b_2}{r_{1}^{10}}\right) 
\end{align*}

\paragraph*{\emph{Expression of} $\Theta_6=[LA_{(0,1)}-LA_{(0,0)},Z_{3}]$.} 
We have:
\begin{align*}
\Theta_6[3,1]
&= 
0\\
\Theta_6[3,2]
&=
m_{01}+(m_{00}-m_{01})E-\frac{\pi}{2} (a_{1}r_2^{00}+a_2r_{1}^{00})\\
&\quad
+\frac{\pi}{2} (c_{1}r_2^{10}+c_2r_{1}^{10})+ \frac{\pi}{2\sqrt{5}}(c_{1}r_2^{10}-c_2r_{1}^{10})\\
\Theta_6[4,2]
&=
-2(m_{00}+m_{01})+\pi (a_2r_{1}^{00}-a_{1}r_2^{00}) 
- \frac{2\pi}{\sqrt{5}}(c_{1}r_2^{10}-c_2r_{1}^{10})\\
\Theta_6[1,3]
&=
\pi \left(\frac{a_2}{r_{1}^{00}}- \frac{a_{1}}{r_2^{00}} \right)-\frac{2\pi}{\sqrt{5}}\left( \frac{c_{1}}{r_2^{10}}-\frac{c_2}{r_{1}^{10}}\right)\\
\Theta_6[1,4]
&=
m_{00}-m_{01} - \frac{\pi}{2} \left( \frac{a_{1}}{r_2^{00}} +\frac{a_2}{r_{1}^{00}} \right)\\
&\quad
+\frac{\pi}{2}\left(\frac{c_{1}}{r_2^{10}}+ \frac{c_2}{r_{1}^{10}}\right)  -\frac{\pi}{2\sqrt{5}}\left( \frac{c_{1}}{r_2^{10}}-\frac{c_2}{r_{1}^{10}}\right)\\
\Theta_6[2,4]
&= 
0.
\end{align*}

It remains to check that these six matrices are linearly independent, at least for all $E>2$ except those in a discrete set. We denote by $f_2(E)$ the determinant of the $6\times 6$ matrix whose columns are representing the $6$ matrices we just compute. Each column is made of the $6$ entries we compute for each matrix. We also set: 
\begin{equation}\label{detbracket2}
f_2(E)= \tilde{f}_2(a_{1},a_2,b_{1},b_2,c_{1},c_2,d_{1},d_2,m_{00},m_{01},m_{10},m_{11} ,E)
\end{equation}
where $\tilde{f}_2(X_{1},\ldots,X_{12},Y)$ is polynomial in the coefficients $X_{1},\ldots,X_{12}$  and analytic in $Y$. 

We define the functions $\tilde{f}_{2,p,l}$ as we defined the functions $\tilde{f}_{1,p,l}$. We can show that the $\tilde{f}_{2,p,l}$ do not vanish identically on $I_{k}$. More precisely we can look at the term in the development of the determinant (\ref{detbracket2}) involving only $a_2$:  
\begin{align*}
&
\frac{m_{10}(m_{11}-m_{10})\pi^{2}a_2^{2}}{4(E+1)^{3}} 
[ \pi a_2m_{11}(10\sqrt{E+1} E^{3}-8(E+1)^{3/2}E^{3}\\[1mm]
&
-9(E+1)^{7/2}+(E+1)^{5/2} +28\sqrt{E+1}E^{2}+14(E+1)^{3/2}E\\[1mm]
&
-2(E+1)^{3/2}E^{2}-11(E+1)^{5/2}E+ 8(E+1)^{7/2}E+26\sqrt{E+1}E\\[1mm]
&
+8(E+1)^{3/2}+8\sqrt{E+1})
+\pi a_2m_{10}(10(E+1)^{5/2}+2(E+1)^{7/2}\\[1mm]
&
+ 8(E+1)^{3/2}E^{3}+ 14(E+1)^{5/2} E-8(E+1)^{7/2} E -29\sqrt{E+1}E^{2}\\[1mm]
&
- (E+1)^{3/2}E+ 10(E+1)^{3/2}E^{2} -28\sqrt{E+1}E- 3(E+1)^{3/2}\\[1mm]
&
-9\sqrt{E+1}-10\sqrt{E+1}E^{3})\\[1mm]
&
+m_{10}m_{11}(16E^{4}+32E^{3}-16E^{2}-64E-32)].
\end{align*}
This term is different from $0$ for $a_2 \geq 1$, $m_{10}\geq 1$, $m_{11}\geq 1$ and $m_{10}\neq m_{11}$. But we can always assume that these two integers are distinct. Indeed, in the proof of Proposition \ref{diop}, we can replace $m_{10}$ by $2m_{10}$ and multiply by $2$ the integers $x_{1}^{10}$ and $x_{1}^{10}$. And of course $m_{10}$ and $2m_{10}$ cannot be both equal to $m_{11}$. 

The term we just computed is the only one in the development of the determinant (\ref{detbracket2}) involving exactly those powers of $E$ and $E+1$ in the numerator and in the denominator. So this term cannot be cancelled uniformly in $E$ by another term of the development of the determinant (\ref{detbracket2}). As before, the functions $\tilde{f}_{2,p,l}$ do not vanish identically on $I_{k}$ whenever $(p,l)\in J_2$ with:
\begin{align*}
J_2
&=
\{(p_1,\dots,p_8,l_1,\dots,l_4)\mid 0\leq p_1,p_3,\dots,p_8\leq N_k, 1\leq p_2\leq N_k,\\
&\qquad\qquad\qquad\qquad\qquad\quad\;\;\, 1\leq l_j\leq M, l_3\neq l_4\}.
\end{align*}
As we have justified that $(a_{1},\ldots ,m_{11})\in J_2$, we have: 
\[
\{E \in I_{k} \ |\ f_2(E)=0\} \subset \bigcup_{(p,l)\in J_2} \{ E\in I_{k} \ |\ \tilde{f}_{2,p,l}(E)=0 \}.
\]
So the set of zeros of $f_2$ is a discrete subset in $(2,+\infty)$. If we set:
\[
\mathcal{S}_{3}=\{E>2\mid f_2(E)=0 \},
\]
$\mathcal{S}_{3}$ is discrete, and for $E>2$, $E\notin\mathcal{S}_{1}\cup\mathcal{S}_2\cup \mathcal{S}_{3})$, 
then $f_2(E)\neq 0$. So for these energies, the matrices 
\begin{alignat*}{3}
&
LA_{(1,0)}-LA_{(0,0)},&\quad&LA_{(1,0)}-LA_{(1,1)},&\quad&LA_{(0,1)}-LA_{(0,0)},\\
&
 [LA_{(1,0)}-LA_{(0,0)},Z_{1}],&&[LA_{(1,0)}-LA_{(1,1)},Z_2],&&[LA_{(0,1)}-LA_{(0,0)},Z_{3}]
\end{alignat*}
are linearly independent in the $6$-dimensional subspace $V_2$. So, 
\[
\text{for all }E>2,\;E\notin\mathcal{S}_{1}\cup \mathcal{S}_2\cup \mathcal{S}_{3},\
\text{ we have }
V_2\subset \mathfrak{la}_2(E).
\]
Finally, we set 
\[
\mathcal{S}_{\mathrm{B}}=\mathcal{S}_{1}\cup \mathcal{S}_2\cup \mathcal{S}_{3}..
\]
We fix $E>2$, $E\notin\mathcal{S}_{\mathrm{B}}$. We have $V_{1} \subset \mathfrak{la}_2(E)$ and  $V_2\subset \mathfrak{la}_2(E)$. As $V_{1}\oplus V_2=\mathfrak{sp}_2(\R)$, we get: 
\[
\text{for all }E>2,\, E\notin\mathcal{S}_{\mathrm{B}},\quad \mathfrak{sp}_2(\R) \subset \mathfrak{la}_2(E)
\]
We have proven: 
\[
\text{for all }E>2,\;E\notin\mathcal{S}_{\mathrm{B}},\quad \mathfrak{sp}_2(\R) = \mathfrak{la}_2(E).
\]

This ends our study of the Lie algebra $\mathfrak{la}_2(E)$. We have proven that for $E>2$, 
$E\notin\mathcal{S}_{\mathrm{B}}$, we can apply Theorem \ref{breuillard} to the four matrices 
\[
A_{0}^{(0,0)}(E)^{m_{00}(E)},\ A_{0}^{(1,0)}(E)^{m_{10}(E)},\ A_{0}^{(0,1)}(E)^{m_{01}(E)},\
A_{0}^{(1,1)}(E)^{m_{11}(E)}.
\] 
Indeed, they are all in $\mathcal{O}$ and their logarithms generate the whole Lie algebra $\mathfrak{sp}_2(\R)$. So this achieves the proof of Proposition \ref{H2zar}.

\subsection{End of the proof of Theorem \ref{HAB2thm}}

We have to explain how we deduce Theorem \ref{HAB2thm} from Proposition \ref{H2zar}. Let $E>2$,
$E\notin\mathcal{S}_{\mathrm{B}}$ be fixed. By Proposition \ref{H2zar}, $G_{\mu_{E}}$ is dense,
therefore Zariski-dense, in $\mathrm{Sp}_2(\R)$. So, applying Theorem \ref{algthm}, we get that $G_{\mu_{E}}$ is $p$-contractive and $L_{p}$-strongly irreducible for all $p$. Then applying Corollary \ref{lyapsepcor} we get the separability of the Lyapunov exponents of the operator $H_{\AB}(\omega)$ and the positivity of the two leading exponents. Thus we obtain Theorem \ref{HAB2thm}: for all $E>2$, $E\notin\mathcal{S}_{\mathrm{B}}$, we have
\[
\gamma_{1}(E)> \gamma_2(E) >0.
\]

\subsection{Proof of Corollary \ref{HAB2cor}}
Corollary \ref{HAB2cor} says that $H_{\AB}(\omega)$ has no absolutely continuous spectrum in $(2,+\infty)$. For this we refer to Kotani's theory in \cite{kotanis}. Note that \cite{kotanis} considers $\R$-ergodic systems, while our model is $\Z$-ergodic. But we can use the suspension method provided in \cite{kirsch} to extend the Kotani's theory to $\Z$-ergodic operators. So, non-vanishing of all Lyapunov exponents for all energies except those in a discrete set allows to show the absence of absolutely continuous spectrum via Theorem~7.2 of \cite{kotanis}. 
\section*{Acknowledgements}
The author would like to thank Anne Boutet de Monvel and G\"unter Stolz for numerous helpful suggestions and remarks, and also for their constant encouragements during this work.

\end{document}